# Perpendicular magnetic anisotropy and Dzyaloshinskii-Moriya interaction at an oxide/ferromagnetic metal interface

Weinan Lin[1†], Baishun Yang[2†], Andy Paul Chen[3], Xiaohan Wu[1], Rui Guo[1], Shaohai Chen[1], Liang Liu[1], Qidong Xie[1], Xinyu Shu[1], Yajuan Hui[1], Gan Moog Chow[1], Yuanping Feng[3,4], Giovanni Carlotti[5], Silvia Tacchi[6*], Hongxin Yang[2*], Jingsheng Chen[1,3*]

1 Department of Materials Science and Engineering, National University of Singapore, 117575, Singapore

2 Ningbo Institute of Materials Technology and Engineering, Chinese Academy of Sciences, Ningbo 315201, China; Center of Materials Science and Optoelectronics Engineering, University of Chinese Academy of Sciences, Beijing 100049, China

3 NUS Graduate School of Integrative Sciences and Engineering, National University of Singapore, 117456, Singapore

4 Department of Physics, National University of Singapore, 117576, Singapore

5 Dipartimento di Fisica e Geologia, Università di Perugia, Via Pascoli, I-06123, Perugia, Italy

6 Istituto Officina dei Materiali del CNR (CNR-IOM), Sede Secondaria di Perugia, c/o Dipartimento di Fisica e Geologia, Università di Perugia, I-06123 Perugia, Italy

[†]These authors contributed equally to this work.

*Corresponding authors: msecj@nus.edu.sg; hongxin.yang@nimte.ac.cn; tacchi@iom.cnr.it

**Abstract:** We report on the study of both perpendicular magnetic anisotropy (PMA) and Dzyaloshinskii-Moriya interaction (DMI) at the oxide/ferromagnetic metal (FM) interface, i.e. $BaTiO_3$ (BTO)/CoFeB. Thanks to the functional properties of the BTO film and the capability to precisely control its growth, we are able to distinguish the dominant role of the oxide termination (TiO$_2$ vs BaO), from the moderate effect of ferroelectric polarization in the BTO film, on the PMA and DMI at the oxide/FM interface. We find that the interfacial magnetic anisotropy energy of the BaO-BTO/CoFeB structure is two times larger than that of the TiO$_2$-BTO/CoFeB, while the DMI of the TiO$_2$-BTO/CoFeB interface is larger. We explain the observed phenomena by first principles calculations, which ascribe them to the different electronic states around the Fermi level at the oxide/ferromagnetic metal interfaces and the different spin-flip process. This study paves the way for further investigation of the PMA and DMI at various oxide/FM structures and thus their applications in the promising field of energy-efficient devices.

Perpendicular magnetic anisotropy (PMA) and Dzyaloshinskii-Moriya interaction (DMI) in conventional ferromagnetic metals (FM) are attracting great interest as they are proposed as key components to design and realize energy-efficient spintronic devices, especially in the recent developed spin-orbit based devices [1,2]. One strategy to enhance the PMA of a conventional FM is to introduce a heavy metal neighbour layer with large spin orbit coupling (SOC) strength at the beginning [3,4], and later evolve to bring in an oxide layer next to it [5,6], which can result in similar strength of PMA. For example, the MgO/CoFeB used in perpendicular magnetic tunnel junctions is promising for realizing the next-generation high-density non-volatile memory and logic chip [5]. On the other hand, the DMI, described by the Dzyaloshinskii-Moriya vector ***D***, is the antisymmetric exchange interaction that promotes canted spin configuration, instead of the parallel or antiparallel spin alignments obtained by usual Heisenberg exchange interaction [7-9]. Though the concept has been introduced several decades ago, only recently it is recognized that DMI can play an important role in electrically manipulation of the magnetization in various materials for achieving potential energy-efficient devices [2,10-12], such as fast domain wall motion [12] and skyrmion lattice formation [13,14]. Similar to PMA, DMI requires the presence of a sizeable SOC, as well as of broken inversion symmetry that is naturally present at interfaces. Therefore, heavy metals, such as Pt and Ir, are usually introduced to engineer the interfacial DMI [13,15] even if it has been shown more recently that an oxide layer can be also exploited to the same aim. This possibility has been analysed from the theoretical point of view in different studies [16,17] even if experimental data are still lacking in the literature. To this respect, a large variation in the interfacial DMI under the application of an electric field has been observed very recently in the Ta/CoFeB/TaO$_x$ system [18]. Oxide materials are versatile, featuring peculiar degrees of freedom, such as the terminations in a complex oxide and the polarization in a ferroelectric oxide, which can be exploited to manipulate the DMI strength. However, a detailed analysis of these effects has been not performed so far.

In this work, we make use of the ferroelectric BaTiO$_3$ (BTO) as an oxide layer to investigate both PMA and DMI at the oxide/FM interface. The precise control of the termination and the polarization of the BTO film helps us to distinguish the role of the termination and polarization in influencing the strength of the PMA and DMI at the oxide/FM interface. With the help of first principles calculations, we ascribe the modulation of the PMA and DMI at the interface to the dominant role of the oxide termination and thus to the different electronic states and spin flip possibilities around the Fermi level.

The studied structures were synthesised in a pulsed laser deposition (PLD)-Sputter combined chamber. The terminations of the BTO film (BaO and TiO$_2$) are realized by controlling the terminations of the SrTiO$_3$ (STO) substrate and the layer-by-layer growth of the BTO film (monitored by the high energy electron diffraction (RHEED) system), i.e. TiO$_2$ terminated STO results in TiO$_2$ terminated BTO, while SrO terminated STO results in BaO terminated BTO. The TiO$_2$ terminated STO is obtained via the conventional buffer-HF solution treatment [19], while the SrO termination is obtained via the deposition of a SrRuO$_3$ (SRO) single layer on top of the treated STO [20]. In this study, 15 unit cell (uc) of BTO are grown on STO substrates of both terminations, as shown in Fig. S1 of the Supplementary Materials. Next, the film is transferred in the sputter chamber for *in situ* growth of the CoFeB film with various thicknesses and then the heavy metal capping layer (Ta or Pt), followed by an annealing procedure (300 °C for 1h). As shown in inset of Fig. 1(a) and (b), the terraced morphology preserved in structures of both terminations. According to our previous works [21,22], besides the different termination of the BTO layers, the different terminated STO substrates will lead to different as-grown polarizations of the BTO layer, i.e. TiO$_2$ (SrO) termination results in the down (up) ferroelectric polarization in the BTO film. In the following, we use the BaO-BTO and TiO$_2$-BTO to indicate these two types of BTO structures, which also hold the up and down ferroelectric polarizations, respectively.

Figure 1(a) and (b) show the normalized MH loops measured with the in-plane (IP) and out-of-plane (OP) applied magnetic fields for a CoFeB film, 1.54 nm thick, grown on TiO$_2$-BTO and BaO-BTO, respectively. Though the growth condition and thickness of CoFeB are identical, it is interesting to note that the CoFeB film grown on TiO$_2$-BTO shows IP magnetic easy axis, whereas for the CoFeB film grown on BaO-BTO an OP easy axis (i.e. perpendicular magnetic anisotropy (PMA)) is observed. To quantify the perpendicular magnetic anisotropy energy (MAE), we estimate the sheet effective magnetic constant ($K_{eff}t_{eff}$) of both samples, which are 0.128 and -0.254 mJ/m$^2$ for CoFeB grown on BaO-BTO and TiO$_2$-BTO films, respectively. As shown in insets of Fig. 1(a) and (b), the similar topographies of both structures exclude the extrinsic factors for the different features.

To investigate the role of the interface on the MAE of both structures [23], we prepared two series of samples with various CoFeB thickness on the TiO$_2$-BTO and BaO-BTO substrates. Figure 1 (c) and (d) plot the magnetic moments per unit area of both terminated structures as functions of the CoFeB thickness. For both structures, the thickness of the magnetic dead layer as well as the saturation magnetizations $M_s$ of the CoFeB films are obtained by linear fittings of $M_{sheet}$ as a function of $t_{CoFeB}$. The thickness of the magnetic dead layer are 0.38 and 0.41 nm for the TiO$_2$-BTO and BaO-BTO structures, respectively, which is mainly resulted from B interdiffusion at the CoFeB/Ta interface [24], while their saturation magnetizations of the CoFeB grown on the TiO$_2$-BTO and BaO-BTO films are 1148 and 1247 emu/cc, respectively.

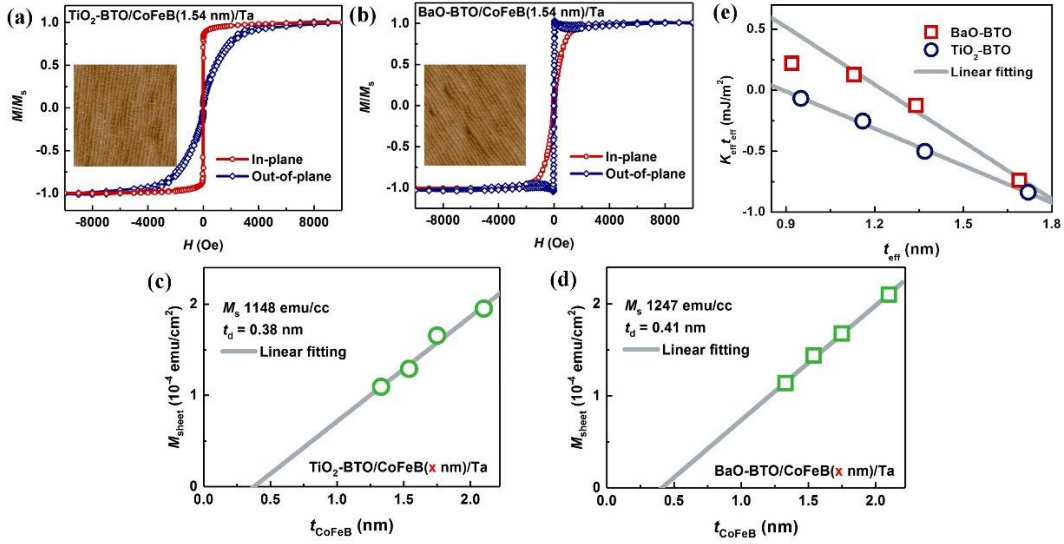

**FIG. 1.** (a) and (b) MH loops measured by SQUID at 300 K for CoFeB (1.54 nm)/Ta bi-layers grown on BaTiO$_3$ with TiO$_2$- and BaO terminations respectively. The insets show AFM topography image of each structure. (c) and (d) Magnetic moment sheet density $M_{sheet}$ as functions of nominal CoFeB thickness for both types of samples. (e) $K_{eff}t_{eff}$ data as functions of the effective thickness of CoFeB for both terminations structures. The grey lines are linear fittings.

Figure 1(e) plots the sheet effective magnetic constant ($K_{eff}t_{eff}$) as functions of the effective CoFeB thickness ($t_{eff}$), where the dead layer thickness is subtracted, for both BaO-BTO/CoFeB and TiO$_2$-BTO/CoFeB structures. As observed from the figure, the effective magnetic constants ($K_{eff}t_{eff}$) for the BaO-BTO/CoFeB structures are always larger than that of the TiO$_2$-BTO/CoFeB structures for all the CoFeB thicknesses. To quantitatively characterize the difference of the magnetic anisotropy of both structures, we fit the data to the equation $K_{eff}t_{eff} = K_i - (2\pi M_s^2 - K_V)t_{eff}$ [23] and obtain the value of the interface anisotropy constant $K_i$ ($K_v$ is bulk magnetic anisotropy). They are 0.9 and 1.93 mJ/m$^2$ for the TiO$_2$-BTO/CoFeB and BaO-BTO/CoFeB structures, respectively, which are at the same magnitude with the standard CoFeB/MgO structure [5]. As mentioned above, besides the different terminations of both structures, their ferroelectric polarizations are also opposite. To distinguish the influence of termination and ferroelectric polarization, we also prepared a controlled structure with 2uc SRO inserted between a 1.33 nm CoFeB and polarized BTO to screen the influence of the ferroelectric polarization [25]. The calculated $K_{eff}t_{eff}$ of the controlled sample is 0.228 mJ/m$^2$, which is close to the value for the CoFeB grown on the BaO-BTO film, 0.222 mJ/m$^2$. Therefore, one concludes that the dominant factor for the substantial difference of the interfacial magnetic anisotropy energy is the oxide termination of the BTO/CoFeB interface.

The interfacial DMI at the oxide/FM interface is investigated by Brillouin light scattering (BLS) [26-29]. In ultrathin films, the presence of i-DMI causes a frequency asymmetry between Damon-Eshbach (DE) modes propagating in opposite in-plane directions, perpendicular to the sample magnetization, following the relation:

$$\Delta f = \frac{2\gamma D}{\pi M_s} k \quad (1)$$

where $D$ is the effective DMI constant, $k$ is the spin waves (SWs) wave vector, $\gamma$ is gyromagnetic ratio. The SW asymmetry, $\Delta f$, is determined by measuring the frequency asymmetry between the Stokes and anti-Stokes peaks in the BLS spectra. To avoid the influence of interdiffusion layer at the FM/HM interface, BLS measurements have been performed for two different CoFeB films, 2 nm thick, grown on a TiO$_2$-BTO or a BaO-BTO substrate, respectively, and capped with a 4 nm thick Pt layer. Figure 2(a) shows the schematic configuration of the BLS experiments from thermally excited SWs, that were carried out by focusing a monochromatic laser beam ($\lambda$ = 532 nm) on the sample surface, and analysing the backscattered light by a Sandercock-type (3+3) pass tandem Fabry-Perot interferometer.

An in-plane magnetic field $H$=3.5 kOe, sufficiently large to saturate the magnetization in the film plane, was applied along the z axis. Meanwhile, the in-plane $k$ was swept along the perpendicular direction (x axis), corresponding to the DE geometry. Due to the conservation of momentum in the light scattering process, the magnitude of $k$ is connected to the incidence angle of light θ, by the relation $k = 4\pi \sin θ/\lambda$. In order to estimate the effective DMI constant D, the SW dispersion (frequency vs wave vector $k$) is measured, changing $k$ from 0 to 2.07 ×10$^7$ rad/m.

Figure 2(b) shows typical BLS spectra measured for the TiO$_2$-BTO/CoFeB and BaO-BTO/CoFeB structures at $k$=1.67×10$^7$ rad/m. The Stokes and anti-Stokes peaks of both samples are characterized by a frequency shift that is more pronounced for the BaO-BTO/CoFeB system. In addition, the frequency of the Stokes and anti-Stokes peaks interchanges on reversing the direction of the applied magnetic field (see Fig. S2 of the Supplementary Materials), that is equivalent, in the interaction geometry sketched in Fig. 2a, to the reversal of the propagation direction of the DE mode.

Figure 2(c) and (d) report the measured (points) frequency asymmetry $\Delta f$, as a function of the wave vector $k$, for the CoFeB/Pt layers grown on $TiO_2$- and BaO-terminated BTO, respectively. $\Delta f$ is obtained as the sum between the measured frequency shift (positive) of the anti-Stokes peak and that (negative) of the Stokes peak. In agreement with Eq. (1), the $\Delta f$ exhibits a linear dependence on the wave vector $k$. The effective DMI constant D has been determined with a linear fit to the experimental data based on Eq. 1. They are 0.45±0.02 and 0.56±0.02 mJ/m$^2$, for $TiO_2$-BTO/CoFeB/Pt and BaO-BTO/CoFeB/Pt structures, respectively. The positive value of D indicates that right-handed chirality is favoured for both systems. Since the CoFeB/Pt interface gives the same contribution to the DMI strength of both systems, the different D values can be attributed to the difference on the BTO terminations.

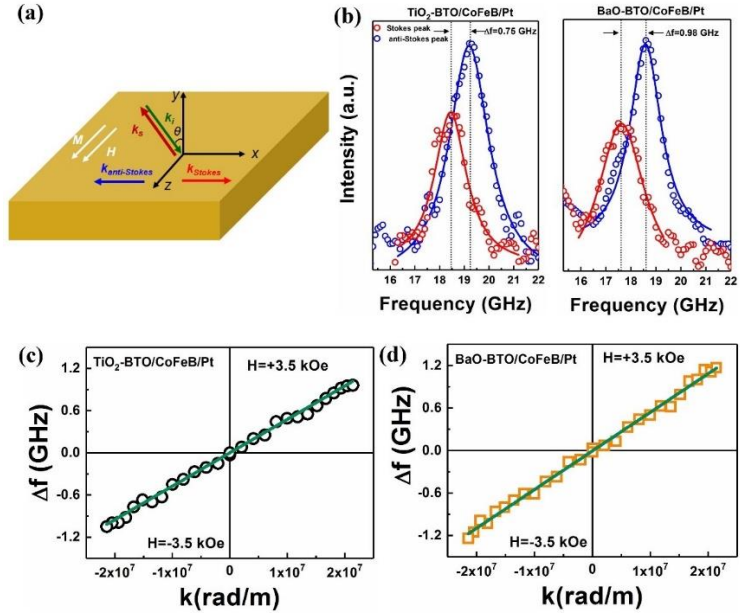

**FIG. 2.** (a) Schematic of Brillouin light scattering (BLS) experiment. (b) The BLS spectra recorded at $k$= 1.67×10$^7$rad/m under an external magnetic field $H$=+3.5 kOe, for CoFeB(2nm)/Pt(4nm) grown on $TiO_2$- (left panel) and BaO- (right panel) terminated BTO. In order to identify the frequency asymmetry between the Stokes (red) and anti-Stokes (blue) peaks, mirror data of the Stokes peak are drawn. Symbols refer to the experimental data and solid lines are the Lorentzian fits. (c) and (d) The measured frequency asymmetry, $\Delta f$, as functions of **k** for CoFeB/Pt grown on $TiO_2$- and BaO- terminated BTO, respectively.

In order to clarify the underlying mechanism of the different PMA and DMI for the two terminated BTO/CoFeB heterostructures, we performed first principles calculations using the Vienna ab-initio simulation package (VASP) [30-34]. The atomic structures of $TiO_2$-termination, with down polarization, and BaO-termination, with up polarization, used in the calculations are shown in Fig. 3(a) and (b), respectively. Since the experimentally synthesized CoFeB is amorphous which is beyond the scope of first principle calculation, and formation energy calculations show that the interfacial Fe-O bonded is energetically more favorable than that of Co-O one [34], therefore we use bcc-Fe instead of CoFeB in the calculations. The thickness of Fe films are 4 monolayer (ML) and 2ML in the calculations of the magnetic anisotropy and DMI, respectively.

**TABLE 1**. Calculated perpendicular magnetic anisotropy energy (MAE) and Dzyaloshinskii-Moriya interaction (DMI) for bilayers of $TiO_2$-terminated BTO/Fe, BaO-terminated BTO/Fe and Fe/Pt as well as trilayers of BTO/Fe/Pt. Positive and negative DMI signs correspond to right-handed and left-handed chirality, respectively.

| Structures | MAE (mJ/m$^2$) | DMI (mJ/m$^2$) |
|---|---|---|
| $TiO_2$-BTO_dn/Fe | 0.38 | -3.85 |
| BaO-BTO_up/Fe | 1.00 | -0.55 |
| Fe/Pt | NA | 8.94 |
| $TiO_2$-BTO_dn /Fe/Pt | NA | 5.09 |
| BaO-BTO_up/Fe/Pt | NA | 8.39 |

The calculated MAE for Fe on BaO-terminated BTO surface is 1.00 mJ/m$^2$, whilst it is only 0.38 mJ/m$^2$ for Fe on $TiO_2$-terminated BTO as shown in Table I. This result is in agreement with the experiment where $K_i$ of the CoFeB film grown on BaO-terminated BTO is nearly twice as large as that of the CoFeB grown on TiO$_2$-terminated BTO. By analyzing the layer-resolved MAE, we find that this difference between the two terminated structures mainly origins from the interfacial Fe layer. Figures 3(c) and (d) show the density of state (DOS) of the interfacial Fe atoms for the two terminated structures. For Fe on BaO-terminatied BTO case, the occupied majority spin states of Fe move downward from the Fermi level compared to Fe on TiO$_2$-terminated BTO case, leading to an appreciably larger spin splitting for Fe in the former case. This also explains stronger saturation magnetization for the CoFeB grown on the BaO-terminated surface compared to the CoFeB grown on the TiO$_2$-terminated one observed in our experiment. On the other hand, in the vicinity of Fermi level, the occupied minority spin $d_{xy}$ state gets much larger in case of Fe on BaO-terminated BTO surface. Figures 3(e) and (f) plot the orbital-resolved MAE of Fe atoms in the two terminated interfaces calculated by the second perturbation theory [34,35]. The largest MAE change comes from ($d_{xy}, d_{x^2-y^2}$) matrix element, where the negative value in TiO$_2$-termination reverses to positive value in BaO-termination, and thus resulting in a larger PMA in BaO-terminated BTO/Fe, which is ascribed to the different orbital hybridizations at the interfaces of the two terminated structures [34, 36].

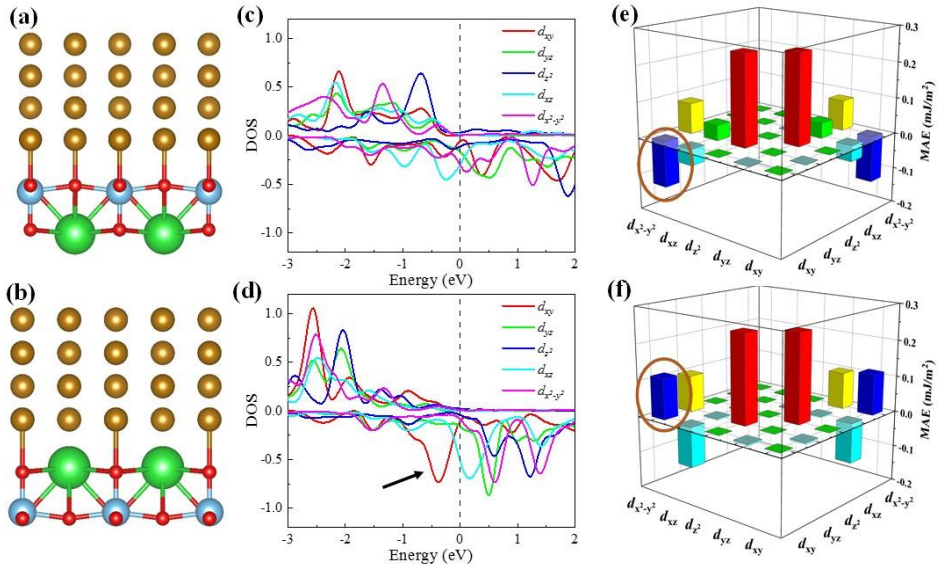

**FIG. 3**. Schematic diagram of BTO/Fe with (a) TiO$_2$-termination and (b) BaO-termination showing down and up ferroelectric polarization, respectively. The density of states of the interfacial Fe atom in (c) TiO$_2$-

and (d) BaO-terminations structures. Orbital-resolved MAE of the interfacial Fe atom at (e) $TiO_2$- and (f) BaO-terminations structures, respectively. The balls represent Fe (yellow), Ba (Green), Ti (blue) and O (red).

Because the DMI mainly comes from the interfaces at FM/HM and FM/Oxide [17, 35], we separated the trilayers structure into two parts, BTO/Fe and Fe/Pt, for the calculation of the DMI, whose schematics are shown in Fig. 4. From the obtained values reported in Table I, one can see that the DMI of Fe/Pt interface dominates the total DMI and shows opposite chirality compared to that of the BTO/Fe interface. Summing up the DMI values of the two interfaces [17], we obtain values of the DMI energy of 5.09 and 8.39 mJ/m$^2$ for the $TiO_2$- and BaO-terminated BTO/Fe/Pt heterostructures, respectively. These results are in qualitative agreement with the experiments, indicating that the total DMI results from the competition between the Fe/Pt and the BTO/Fe interfaces. The larger calculated DMI values may come from the different thickness of the FM layer in experiments and calculations, since the micromagnetic DMI is inversely proportional to the thickness of magnetic films [37]. Concerning the difference of DMI in the two terminated surfaces, one should consider, that besides the spin orbital coupling in inversion symmetry broken system, the band-filling of the 3$d$ atom plays an important role in determining the DMI strength in conventional 3$d$/5$d$ interfaces [38]. Though without heavy metals in these two terminated interfaces, the band-filling of the Fe layer may still dominate the differences of their DMI strengths [39]. As seen in Fig. 3(d), the interfacial Fe atom in BaO-termination shows large spin splitting, i.e. there are fewer electrons with spin-up around the Fermi level, so that the electronic states around the Fermi level show the same spins and hence the spin-flip processes are less favored than that in the $TiO_2$-terminatied structure, which has more occupied spin-up states and vacant spin-down states around the Fermi level simultaneously. As a result, the DMI of BaO-terminated BTO/Fe is smaller than that of the $TiO_2$-terminated one.

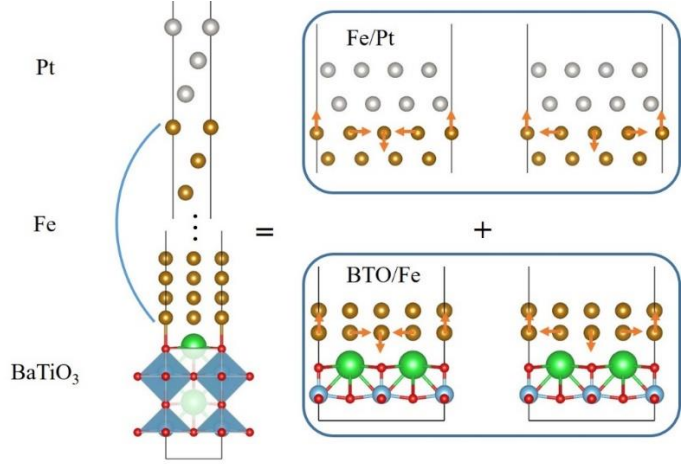

**FIG. 4**. Schematic for calculation of the DMI of the trilayers, which is separated into two parts from the upper and bottom interfaces. The DMI at each interface is calculated as the energy difference of the structure right-handed and left-handed chirality (inside boxes). (Only the BaO-terminated structure shown here).

Finally, in order to gain a deeper insight into the role of ferroelectric polarizations on the observed differences of the PMA and DMI, we have performed first principles calculations for the $TiO_2$-BTO/Fe structure with up polarization. The calculated MAE increases by 6% with respect to the down polarization, but remains far smaller than that of the BaO terminated structure. Moreover, the calculated DMI of $TiO_2$-BTO/Fe with up polarization turns out to reduce its magnitude by about one third, but it is still of the same order of magnitude of $TiO_2$-BTO/Fe with down polarization and is much larger than that of the BaO-BTO/Fe structure. The potential atomic relaxation effect on the PMA and DMI has been considered [34], which is moderate compared to the effect from the terminations. Therefore, one concludes that the termination of the BTO layer, rather than the ferroelectric polarization, plays a dominant role in the modulation of the PMA and DMI in the investigated BTO/CoFeB structures.

In summary, the effects of changing oxide termination and ferroelectric polarization on PMA and DMI at the BTO/CoFeB interface have been investigated. We found that the choice of the termination strongly affects both PMA and DMI strength. In particular, a larger PMA has been observed for the CoFeB films grown on a BaO-BTO substrate, while a higher value of the DMI constant has been found for a $TiO_2$-BTO substrate. First principle calculations show that this behaviour can be ascribed to the different electronic states around the Fermi level at the oxide/FM interfaces. This provides another degree of freedom to manipulate the PMA and DMI in a FM layer. These results may inspire further studies of the interface characteristics in various oxide/FM systems, paving the way to the design of layered structures with tailored DMI to be exploited in forthcoming energy-efficient devices.

**Acknowledgements**

The research is supported by the Singapore National Research Foundation under CRP Award No. NRF-CRP10-2012-02 and Singapore Ministry of Education MOE2018-T2-2-043, AMEIRG18-0022, A*STAR IAF-ICP 11801E0036 and MOE Tier 1- FY2018–P23. Financial supports by the National Natural Science Foundation of China (11874059), Zhejiang Province Natural Science Foundation of China (LR19A040002), the European Metrology Programme for Innovation and Research (EMPIR), under the Grant Agreement 17FUN08 TOPS, are kindly acknowledged. G. M. C would like to acknowledge the financial support from the Singapore Ministry of Education (MOE2018-T2-1-019 and MOE T1 R-284-000-196-114).